\documentclass[aps,pra,showpacs]{revtex4}
\usepackage{graphicx}
\begin{document}
\title{Surface Screening in the Casimir Force }
\author{Ana M. Contreras-Reyes}
\affiliation{ Centro de Ciencias F\'isicas, Universidad Nacional Aut\'onoma
de M\'exico, Avenida Universidad S/N, Cuernavaca, Morelos 62210, M\'exico}
\author{W. Luis Moch\'an}
\affiliation{ Centro de Ciencias F\'isicas, Universidad Nacional Aut\'onoma
de M\'exico, Avenida Universidad S/N, Cuernavaca, Morelos 62210, M\'exico}
\begin{abstract}
We calculate the corrections to the Casimir force between two metals
due to the 
spatial dispersion of their response functions. We employ
model-independent expressions for the force in terms of the optical
coefficients. We express the non-local corrections
to the Fresnel coefficients employing the surface $d_\perp$ parameter,
which accounts for the distribution of the surface screening
charge. Within a self-consistent jellium calculation, spatial
dispersion increases the Casimir force significatively for small
separations. The nonlocal correction has the opposite sign than
previously predicted employing hydrodynamic models
and assuming abruptly terminated surfaces.
\end{abstract}
\maketitle

The Casimir force between two ideal mirrors a small distance apart is a
manifestation of the quantum fluctuations of
the electromagnetic field confined to the cavity within \cite{Casimir}.
The study of vacuum forces among real materials began with Lifshitz
\cite{Lifshitz} in
1956, who obtained a formula for the force between semi-infinite
homogeneous isotropic local systems characterized by a frequency $\omega$
dependent dielectric function $\epsilon(\omega)$. 
Developments in micro- and nano-technologies have stimulated a renewed
attention on the Casimir force. Deviations from the force between
ideal mirrors are now measurable experimentally. The pioneering
measurements of Lamoreux \cite{Lamoreaux} had
a $5\%$ precision and $1\%$ precision has
now become common \cite{Mohideen, Roy, Harris,
Decca}. Distances down to $\approx 60$nm have been explored \cite{PRA69}. 
Due to alignment difficulties, most experiments have involved
spherical surfaces, while most theories have concentrated on the force
between flat 
parallel surfaces for simplicity, so comparisons have resorted to the 
{\em proximity theorem} \cite{TFP} which relates approximately the
force between curved surfaces to that between flat parallel
surfaces. Nevertheless, measurements of the force
between flat surfaces have already been performed \cite{pparalelas}. 
We expect that in the near
future even better measurements at smaller distances will be
produced. Therefore, manifold 
corrections have to be carefully evaluated in theoretical
calculations, such as the finite conductivity of the metals
\cite{conduct1,conduct2,conduct3,Lambrecht}, surface
roughness \cite{rugoso1,rugoso2}, finite
temperatures \cite{thermal1,thermal2}, grain structure \cite{PRA69}, etc.
In this paper we calculate non-local effects in the Casimir force. We
employ density functional jellium calculations \cite{Feibelman,Ansgar}
of the surface response functions which are consistent with realistic
electronic density profiles that decay smoothly to zero beyond the
nominal surface of the metal, instead of being artificially truncated
abruptly \cite{rcl,heinrichs75}.  At very small distances the
non-local corrections can be 
remarkably large. Within simple hydrodynamic models with abrupt
surfaces \cite{rcl,heinrichs75} nonlocal effects yield a
diminution of the force apparently due to the possibility of exciting
bulk plasmons at the surface of spatially dispersive metals. However, 
nonlocal corrections within self-consistent jellium  theories have the
opposite sign and yield an increase in the Casimir force, as most of
the surface screeening charge accumulates outside of the
body of the conductor, as characterized with $d_\perp$ theory
\cite{Feibelman,Ansgar}.

It has been shown \cite{Mochan,rcl} that Lifshitz formula \cite{Lifshitz},
when written in terms of the {\em exact} surface impedances
\cite{stratton,halevi} of the system,
or equivalently, in terms of its optical reflection amplitudes \cite{Reynaud},
\begin{equation}\label{fCasimir}
\frac{F(L)}{A}=\frac{\hbar c}{2 \pi^2}\int_0^{\infty} dQ Q \int_{q>0} d k
\frac{k^2}{q} f \mbox{Re}\left[ \frac{1}{\xi_s-1} +
\frac{1}{\xi_p-1} \right],
\end{equation}
is applicable to any system with translational invariance along the surfaces
and isotropy around their normal \cite{Mochan,rcl}. Here, $F(L)/A$ is
the force per unit area between two 
slabs with flat parallel faces a distance $L$ apart, $c$ the 
speed of light, $q=\omega/c$ the wavenumber, $\vec Q$ the projection
of the wavevector parallel to the surface and $k=\sqrt{q^2-Q^2}$ its
component normal to the surface,
$\xi_{\alpha}=(r^1_{\alpha}r^2_{\alpha}
e^{2i\tilde{k}L})^{-1}$,
$r^a_{\alpha}$ the
reflection amplitude of slab $a=1,2$ corresponding to the polarization
$\alpha=s,p$, 
and $f=\coth(\beta\hbar\omega/2)/2$ the photon
occupation number at temperature $T=1/k_B\beta$. The integral over $k$
in Eq. (\ref{fCasimir}) goes from $iQ$ towards 0 along the imaginary
axis and then towards $\infty$ just above the positive real axis, although
it may be easily manipulated into a more convenient path along the
imaginary axis. 

Nonlocal effects
\cite{halevi} in the Casimir  force have been calculated
approximately \cite{Kats} employing 
expressions for the surface impedance \cite{conduct3} which are only valid
for good conductors at low frequencies \cite{Landau}, and they have
been calculated from the density response function of semiinfinite
metals within a semiclassical inifinite barrier model
\cite{heinrichs75}. However, they 
may be fully incorporated simply by
plugging the appropriate reflection amplitudes into Eq. (\ref{fCasimir}).
The polarization induced at a given position $\vec r$ within a
non-local or spatially dispersive medium depends on the field that is
applied at nearby positions $\vec r'$ \cite{halevi, rojo}, instead
of depending exclusively on the field at $\vec r$. Within the bulk of
a homogeneous system, non-locality manifests itself as a dependence of
the response functions on the wavevector besides their usual
dependence on the frequency. 
Thus, spatial
dispersion is expected to become important as the lengthscales
associated to the field become small. A few consequences of spatial
dispersion at metal surfaces, such as the anomalous skin effect
\cite{skin}, the excitation at surfaces of propagating plasmons
\cite{rcl}, and the finite screening distance \cite{heinrichs75} of
low frequency fields, 
have been studied in relation to the Casimir force, which they modify
substantially at very small distances.

The most simple non-local model for the response of a metal is the
hydrodynamic model \cite{forstmann,MHidro}, in which a negative semiclassical
electronic 
fluid is confined within a uniform  positively charged
region. The electrons interact with the macroscopic electric field and
with the 
hydrodynamic forces derived from the pressure of the fluid.  The longitudinal
contribution to the bulk dielectric response in this model is
\begin{equation}\label{epsilonL}
\epsilon^L((\vec Q,k_l), \omega)=  1 - \frac{\omega_p^2} {\omega^2 +
i\omega/\tau - \beta^2(Q^2+k_l^2)},
\end{equation}
where the compressibility, related to $\beta ^2=3v_F^2/5$, is the
source of the spatial dispersion. Here, $v_F$ is Fermi's velocity,
$\omega_p$ the plasma frequency,  $\tau$ a phenomenological lifetime,
and $k_l$ is the component of the wavevector of the longitudinal wave
normal to the surface, given by the longitudinal dispersion relation
$\epsilon^L=0$. In this model, the reflection amplitude $r_s$
for $s$ polarized light agrees with that, $r_s^0$, of a local metal
described by 
the Drude dielectric funcion
$\epsilon^T(\omega)=1-{\omega_p^2}/({\omega^2+i\omega/\tau})$.
To calculate the reflection amplitude $r_p$ for $p$ polarized waves,
the excitation of longitudinal waves at the surface has to be
accounted for. This might be done by postulating the continuity of the
component $E_\perp$ of the electric field normal to the
surface as a physically
reasonable additional boundary condition (meaning there is no
singularity in the induced charge density), beyond the usual conditions
derived from electromagnetic theory, and solving them for the
amplitudes of the longitudinal transmitted field as well as the
transmitted and reflected transverse waves, yielding
\begin{equation}\label{rplong}
r_p=\frac{\epsilon^T k - k_m + Q^2(\epsilon^T - 1)/k_l}
{\epsilon^T k + k_m - Q^2(\epsilon^T - 1)/k_l},
\end{equation}
where $k_m=\sqrt{\epsilon^T q^2-Q^2}$ is the normal component of the
wavevector of the transverse wave transmitted into the metal
\cite{error}. The reflection amplitude $r_p^0$ for a local metal may
be obtained from Eq. (\ref{rplong}) by taking the limit $\beta\to 0$,
$k_l\to\infty$. 

In order to use more realistic models of metallic surfaces, we require
expressions for the reflection amplitudes that may also be employed beyond the
hydrodynamic model. We remark that non-local
effects are expected to be important mostly within a thin region close
to the surface of the metal, where the electric field varies rapidly
from its value in vacuum to its bulk value. As  the width
of the selvedge is typically much smaller than the wavelength
$\lambda$, $r_p$ may be calculated
perturbatively\cite{mochan83}. Applying boundary conditions across the
selvedge and employing the long wavelength approximation we obtain
\cite{Feibelman,Ansgar}
\begin{equation}\label{rpr}
r_p=r_p^0 \left[1+\frac{2ik \epsilon^T} {\epsilon^T
k^2/Q^2+1} d_{\perp} \right],
\end{equation}
where $d_\perp/\lambda$ is the small
perturbative parameter and
\begin{equation}\label{dperp}
d_{\perp}(\omega)\equiv \frac{\int dz \ z \delta\rho(z)}{\int dz
\delta\rho(z)},
\end{equation}
is the position of the centroid of the induced charge density $\delta\rho(z)$ 
measured towards vacuum from the nominal surface of the metal
\cite{Feibelman}, 
defined by the edge of the positive background. In a local model with
an abrupt interface, there is a singular charge induced right at the
surface, so $d_\perp=0$. In non-local models with an abrupt
termination the screening charge occupies a finite region within the
metal, so that $d_\perp<0$, as may be confirmed within the
hydrodynamic model, for
which 
\begin{equation}\label{dhydro}
d_{\perp}=-\frac{i}{k_l}.
\end{equation}
In this case, $-d_\perp$ may be regarded as the finite frequency
generalization of the Thomas-Fermi screening distance. However,
within more sophisticated models such as the self 
consistent jellium model \cite{Feibelman,Ansgar} with a smoothly
terminated charge density profile, the screening charge may be
localized on the outside of the surface, so $d_\perp$ may
become positive. Furthermore, surface dissipation yields a complex
valued $d_\perp$.

Linearizing the Casimir force (\ref{fCasimir}) with respect to
$d_\perp$ , we 
obtain an expression for the non-local correction
\begin{equation}\label{Frvsd}
\frac{\delta F(L)}{A} =\frac{2\hbar c}{ \pi^2}\int_0^{\infty} dQ Q
\int_{q>0} dk \frac{k^2}{q} f \mbox{Im} \left[
\left( \frac{ \epsilon^T k d_{\perp}}{\epsilon^T k^2/Q^2-1}
\right) \frac{\xi_p^0}{(\xi_p^0-1)^2} \right],
\end{equation}
where $\xi^0_p$ is the local limit of $\xi_p$. This expression may be
simplified at small distances for which retardation may be neglected,
yielding
\begin{equation}\label{nonret}
\frac{\delta F(L)}{A} =\frac{2\hbar c}{ \pi^2}\int_0^{\infty} dQ Q^3
\int_0^\infty dq f \mbox{Im} \left[\left( \frac{\epsilon^T
d_{\perp}}{\epsilon^T+1} \right) \frac{\xi_p^0}{(\xi_p^0-1)^2} \right],
\end{equation}
where we may further approximate $\xi_p^0 = [(\epsilon+1)/(\epsilon-1)]^2
e^{2QL}$. 

\begin{figure}
\begin{center}
\scalebox{1}{%
%\scalebox{.65}{%
%\input{Hvsd}%
\input{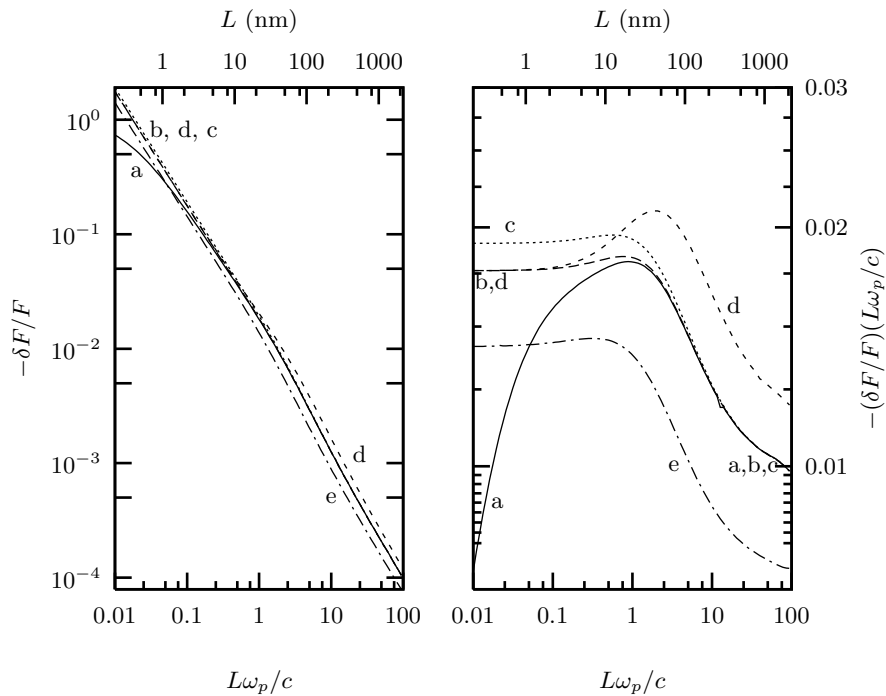}%
}%
\end{center}
\caption{\label{Hvsd} Nonlocal correction to the Casimr 
force between Au flat surfaces  (a-d) and between a flat and a
curved surface (e), calculated within the hydrodynamic model using the
exact (a) reflection amplitudes, retarded
$d$-parameter theory using the dynamical values of $d_\perp(\omega)$
(b, e) and
its static limit $d_\perp(0)$ (c), and the non-retarded $d$-parameter
theory (d) as a function of the separation $L$ between plates. The
left panel shows the normalized correction $-\delta F/F$ and the right
panel shows the correction $(-\delta F/F)(L\omega_p/c)$ scaled by the
separation $L$.}
\end{figure}

Fig. (\ref{Hvsd}), calculated within the hydrodynamic model using the
density parameter 
$r^s=1.59$\AA\ and the lifetime $\tau=400/\omega_p$ corresponding to
Au\cite{solido}, shows that the non-local correction decreases the Casimir
force within the hydrodynamic model.  At
$L=60nm\approx3c/\omega_p$, the closest 
distance for which measurements are available, the non-local
correction amounts to $\delta F/F\approx 0.5\%$, which is smaller \cite{error}
than the experimental error estimate of $1.75\%$ \cite{PRA69}.
However, $\delta F/F$ increases for shorter distances approximately as
$L^{-1}$ and may become of the order of 100\% at a few \AA. 
Although non-local corrections seem
too small to be noticeable in current experiments, they will become
much more important at closer range.

We use now the hydrodynamic model as a test ground to appraise the
results of the perturbative theory. In order to enhance the differences
between the different
approximations discussed above, in
the right panel of Fig. \ref{Hvsd} we show the
corresponding nonlocal corrections scaled by the separation $L$.
For distances larger than $L\approx2nm\approx0.1c/\omega_p$ there is
a very good agreement between the perturbative and the exact
hydrodynamic calculations; for that distance perturbative
calculation produces a $\approx 12\%$ error on the  $\approx 15\%$
nonlocal correction. We also show
in Fig. \ref{Hvsd} %and  \ref{Hvsd1} 
that the corrections to the
force between a 
plane and a sphere \cite{PRA69} are
similar to those for flat surfaces. 

The non retarded calculation also yields
similarly good results up to distances $L\approx 20nm\approx
c/\omega_p$, and although it overestimates 
the corrections by up to $\approx 20\%$ at larger distances
$L\approx 10c/\omega_p$, they are themselves
negligible $\delta F/F<0.1\%$ at those distances. 
Finally, we calculated the nonlocal corrections employing the
perturbative theory but substituting the static value $d_\perp(0)$ instead
of the dynamic value $d_\perp(\omega)$. In this case, the nonlocal
correction is further overestimated but by not more than 20\% if
$L>0.1c/\omega_p$. 

As the vacuum fluctuations of the electromagnetic field within a
cavity are correlated to fluctuations in the charges and currents
induced at the surfaces, the $p$-polarization contribution to the
Casimir force may be interpreted as due to the force between
fluctuating charges in one surface and their images charges at the
opposite surface. Within nonlocal optics, the charges induced at a
surface are not at the nominal surface, but are actually spread over a
finite region whose centroid lies a distance $d_\perp$ from the
nominal surface. At low frequencies, the position of the centroid
corresponds to the 
effective {\em image plane} \cite{Feibelman}. Thus, we may expect that
the force 
between nonlocal
media is similar to the force between local media but separated by an
{\em effective} distance $L'=L+\delta L$. As  $F\propto L^{-3}$ at
small distances, $\delta F/F\approx -3\delta L/L$, which allows to
identify $\delta L\approx -0.006 c/\omega_p\approx -2 d_\perp(0)$ from
Fig. \ref{Hvsd}. Thus, 
nonlocal effects amount to displacing each surface a distance
$\sim d_\perp$ towards each other. Since in the hydrodynamic model
$d_\perp<0$, the effective separation is larger than the nominal
separation and nonlocality decreases the force.

Although very simple and amenable to analytical solutions, the
hydrodynamic model for an abruptly terminated electron gas is not
considered a realistic model of metal surfaces. It doesn't account for
the fact that electrons spill across the nominal surface, giving rise
to a static surface dipole which is the source of the confinement
potential, nor 
the density oscillations due to the quantum interference between the
incoming and outgoing wavefunctions of electrons reflected at the
surface. Landau damping due to excitation of electron-hole pairs and
many body effects such as exchange and correlation are also
absent. Nevertheless, much more sophisticated models of metal surfaces
have been developed \cite{Feibelman,Ansgar}, and have been
employed to calculate $d_\perp$ for manifold metals. In the {\em
jellium} model, the Schr\"odinger-like Kohn-Sham equations
\cite{hks} of density functional theory are solved within the local
density approximation (LDA)
to obtain the selfconsistent 
wavefunctions and ground state density of electrons which interact
with an abruptly terminated semiinfinite homogeneous positive
background. Their response to a perturbing electric field may
then be obtained using time dependent LDA, and $d_\perp$ may be
calculated from the ensuing 
density-density response function \cite{Ansgar} using Eq. (\ref{dperp}).

\begin{figure}
\begin{center}
\scalebox{1}{%
%\scalebox{.65}{%
\input{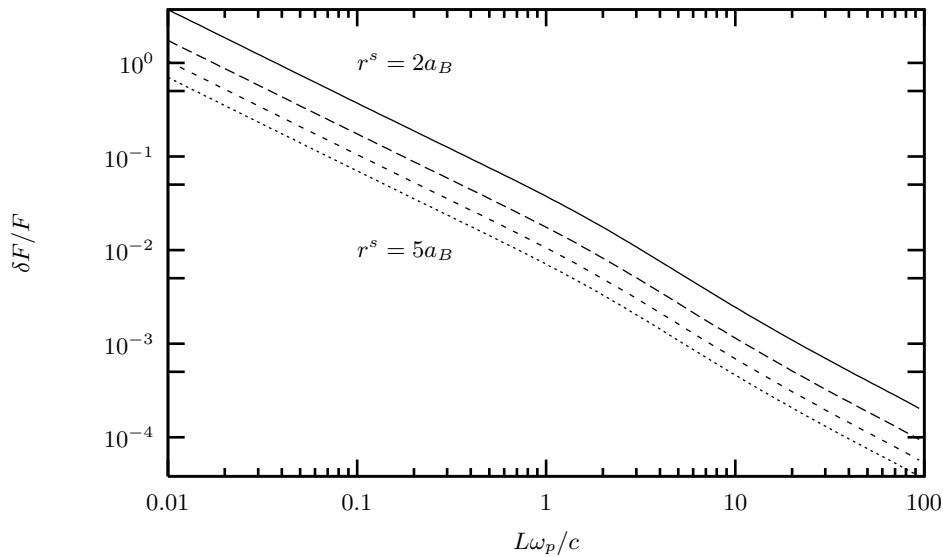}%
}%
\end{center}
\caption{\label{dF_F_allrs} Normalized nonlocal corrections to the
Casimir force calculated for the selfconsistent semiinfinite jellium
model using the LDA for different densities
corresponding to $r^s/a_B=2$, 3, 4, and 5.}
\end{figure}

In Fig. (\ref{dF_F_allrs}) we show the nonlocal corrections to the
Casimir force for metals
of various density parameters $r^s$ \cite{solido} calculated within
the self consistent jellium model obtained by plugging the
corresponding known values of 
$d_\perp(0)$ \cite{Ansgar} into Eq. (\ref{Frvsd}).
In contrast to the previous hydrodynamic results, in this model the nonlocal
effects {\em increase} the Casimir force since most of the screening
takes place in the low density region on the {\em outside} of the nominal
surface, i.e., $d_\perp >0$ and the effective distance $L'$ is smaller
than the nominal distance $L$. We remark that in the
hydrodynamic model the ground state electronic density is constant up
to the surface, where it is abruptly truncated, while in the jellium
model it has a tail that extends beyond the surface. 
The stiffness of the tail of the electronic fluid is smaller than
within the metal, so that it is more easily polarized than the
bulk. Therefore, the
centroid of the induced charge within the jellium model lies outside
of the metal. In the 
hydrodynamic model there is no charge at all out of the metal, so the
centroid of the induced charge is necessarily within the metal and has
the opposite sign.
According to Fig. \ref{dF_F_allrs}, the nonlocal corrections are
larger for 
high density metals (small $r^s$) than for low density metals (high
$r^s$). 

In summary, we have calculated the nonlocal corrections to the Casimir
force between metals using a generalization of Lifshitz formula in
terms of reflection amplitudes which has been proved to be applicable
to arbitrary materials that are homogeneous and isotropic along the
surface. A perturbative 
theory in which the width of the selvedge divided by the relevant
wavelengths plays the role of small parameter is a useful
approximation for distances that are not too small, and yields an
interpretation of the nonlocal effect in terms of an effective
separation between the metal surfaces. 
The exactly solvable
hydrodynamic model 
predicts a diminution in the force as the effective
separation increases due to the finite screening length of spatially
dispersive metals. Contrariwise, the jellium model yields an increase
of the Casimir force 
due to the spilling of electrons out of the metals and into
vacuum, and to the fact that most of the screening takes place in the
resulting low density region outside of the nominal surfaces, thus
reducing the effective separation. 
$|\delta F/F|$ grows roughly as $L^{-1}$ and thus becomes very important
at distances smaller than $\sim 100$nm. As the surface screening may
be manipulated through the adsorption of overlayers \cite{Ansgar}, our results
suggest that the Casimir force at small distances may be
tailored. 
Other surface corrections, such as the surface local field
effect \cite{mochan85} may easily be incorporated into Casimir force
calculations simply by replacing the Fresnel coefficients
$r_p$ and $r_s$ in Eq. (\ref{fCasimir}) by the appropriate nonlocal,
surface-corrected expressions
\cite{mochan83} instead of Eqs. (\ref{rplong}) or (\ref{rpr}).
We hope our results stimulate experiments exploring such small distances.

We acknowledge useful discussions with Carlos Villarreal, Raul
Esquivel, Gregorio H.-Cocoletzi, Catalina L\'opez-Bastidas, and
Claudia Eberlein. This
work was partially supported by DGAPA-UNAM under grant IN117402.


\begin{thebibliography}{00}
\bibitem{Casimir} H. B. G. Casimir, Proc. Kon. Nederl. Akad. Wet {\bf 51}, 793 
(1948)
\bibitem{Lifshitz} E. M. Lifshitz, Zh. Eksp. Teor. Fiz. {\bf 29}, 94 
(1955)[Sov. Phys. JETP {\bf 2}, 73 (1956)].
\bibitem{Lamoreaux} S. K. Lamoreaux, 
%{\em Demonstration of the Casimir Force in the $0.6$ to $6 \mu m$
%range}, 
Phys. Rev. Lett. {\bf 78}, 5 (1997).
\bibitem{Mohideen} U. Mohideen and Anushree Roy, 
%{\em Precision Measurement of the Casimir Force from $0.1$ to $.9 \mu
%m$}, 
Phys. Rev. Lett. {\bf 81}, 4549 (1998); 
G. L. Klimchitskaya, A. Roy, U. Mohideen, and V. M. Mostepanenko, 
Phys. Rev. A {\bf 60}, 3487 (1999).
\bibitem{Roy} A. Roy, C. Y. Lin, and U. Mohideen, Phys. Rev. D {\bf 60}, 
111101(R) (1999).
\bibitem{Harris} B. W. Harris, F. Chen, and U. Mohideen, 
%{\em Precision Measurement of the Casimir force using fold surfaces}, 
Phys. Rev. A {\bf 62}, 052109 (2000).
\bibitem{Decca} R. S. Decca, E. Fischbach, G. L. Klimchitskaya,
D. E. Krause, D.
L\'opez, and V. M. Mostepanenko, Phys. Rev. D {\bf 68}, 116003 (2003).
\bibitem{PRA69} F. Chen, G. L. Klimchitskaya, U. Mohideen, and
V. M. Mostepanenko, 
%{\em Theory confronts experiment in the Casimir force measurements:
%Quantification of errors and precision}, 
Phys. Rev. A {\bf 69}, 022117 (2004).
\bibitem{TFP} J. Blocki, J. Randrup, W. J. Swiatecki, and C. F. Tsang,
Ann. Phys. (N.Y.) {\bf 105}, 427 (1977).
\bibitem{pparalelas} G. Bressi, G. Carugno, R. Onofrio, and G. Ruoso,
%{\em Measurement of the Casimir Force between parallel metallic surfaces},
Phys. Rev. Lett. {\bf 88}, 041804 (2002).
\bibitem{conduct1} C. H. Hargreaves, Proc. K. Ned. Akad. Wet. B {\bf 68}, 231 
(1965).
\bibitem{conduct2} J. Schwinger, L. L. DeRaad, Jr., and K. A. Milton, Ann. 
Phys. (N.Y.) {\bf 115}, 1 (1978).
\bibitem{conduct3} V. M. Mostepanenko and N. N. Trunov, Yad. Fiz. {\bf 42}, 
1297 (1985)[Sov. J. Nucl. Phys. {\bf 42}, 818 (1985)].
\bibitem{Lambrecht} A. Lambrecht and S. Reynaud,
%{\em Casimir force between metallic mirrors}, 
Eur. J. Phys. D. {\bf 8}, 309-318 (2000).
\bibitem{rugoso1} A. A. Maradudin and P. Mazur, Phys. Rev. B {\bf 22}, 1677 
(1980).
\bibitem{rugoso2} M. Bordag, G. L. Klimchitskaya, and V. M Mostepanenko, Int. 
J. Mod. Phys. A {\bf 10}, 2661 (1995).
\bibitem{thermal1} L. S. Brown and G. J. Maclay, Phys. Rev. {\bf 184}, 1272 
(1969).
\bibitem{thermal2} F. Chen, G. L. Klimchitskaya, U. Mohideen, and V. M. 
Mostepanenko, Phys. Rev. Lett. {\bf 90}, 160404 (2003).
\bibitem{Feibelman} P. J. Feibelman, {\em Surface Electromagnetic
fields}, Prog. Surf. Sci. {\bf 12} (1982).
\bibitem{Ansgar} Ansgar Liebsch, {\em Electronic Excitations at Metal
Surfaces} (Plenum, New York, 1997).
\bibitem{rcl} R. Esquivel, C. Villarreal, W. Luis Moch\'an, 
%{\em Exact surface impedance formulation of the Casimir force:
%Application to spatially dispersive metals},
Phys. Rev. A {\bf 68}, 052103 (2003).
\bibitem{heinrichs75}J. Heinrichs,
%{\em Theory of van der Walls interactions between metal surfaces},
Phys. Rev. A {\bf 11}, 3625 (1975).
\bibitem{Mochan} W. Luis Moch\'an, C. Villarreal, and
R. Esquivel-Sirvent, Rev. Mex. Fis. {\bf 48}, 339 (2002).
\bibitem{stratton}J. A. Stratton, {\em Electromagnetic Theory}
(McGraw-Hill, N.Y., 1941).
\bibitem{halevi}P. Halevi, {\em Photonic Probes of Surfaces} (Elsevier,
Amsterdam, 1995); {\em Spatial Dispersion in Solids and Plasmas}
Electromagnetic Waves Vol. 1 (North-Holland, Amsterdam, 1992).
\bibitem{Reynaud} M. T. Jaekel and S. Reynaud, J. Phys. I {\bf 1} 1395
(1991); C. Genet, A. Lambrecht, and S. Reynaud, Phys. Rev. A {\bf 67},
043811 (2003).
\bibitem{Kats} E. I. Kats, Zh. Eksp. Teor. Fiz. {\bf 73}, 212
(1977)[Sov. Phys. JETP {\bf 46}, 109 (1977)].
\bibitem{Landau}L. D. Landau, E. M. Lifshitz, and L. P. Pitaevski\u\i,
{\em Electrodynamics of Continuous Media}, 2nd edition (Pergamon, N. Y., 1984).
\bibitem{rojo} Yu. A. Il'inskii, L. V. Keldysh, {\em Electromagnetic
Response of Material Media}, (Plenum Press, New York, 1994).
\bibitem{skin} R. Esquivel, V. B. Svetovoy,
%{\em Correction to the Casimir force due to the anomalous skin
%effect}, 
Phys. Rev. A {\bf 69}, 062102 (2004).
\bibitem{forstmann} F. Forstmann and R. R. Gerhardts, {\em Metal
Optics Near the Plasma Frequency} (Springer, Berlin, 1986).
\bibitem{MHidro} M. G. Cotram and D. R. Tilley, 
{\em Introduction to Surface and Superlattice Excitations} (Cambridge
University Press, Cambridge, 1989).
\bibitem{error} Notice that that Eq. (14) of Ref. \cite{rcl} is
wrong. Furthermore, the horizontal axis in Fig. 2 of Ref. \cite{rcl}
is mislabeled and off by a factor of $2\pi$; it should have been
$L\omega_p/c$. Additionally, the results for Au in Fig. 2 of
Ref. \cite{rcl} were calculated with the parameter
$\beta=\sqrt{1/3}v_F$ instead of $\sqrt{3/5}v_F$ as in the present
manuscript. See Raul Esquivel, Carlos Villarreal, and W. Luis
Moch\'an, {\em Erratum: Exact surface impedance formulation of the
Casimir force: application to spatially dispersive metals
[Phys. Rev. A {\bf 68}, 052103 (2003)]}, Phys. Rev. A.{\bf 71}, 029904
(2005).
\bibitem{mochan83}%
%Surface Contribution to the Optical Properties of Non-Local Systems,
W. Luis Moch\'an, Ronald Fuchs, and Rub\'en G. Barrera,
Phys. Rev. B {\bf 27}, 771-780  (1983).
\bibitem{solido} Neil W. Ashcroft, N. David Mermin, {\em Solid State
Physics} (Hot-Saunders International Editions, 1976).
\bibitem{hks}P. Hohenberg and W. Kohn, Phys. Rev. {\bf 136}, B864
(1964);
W. Kohn and L. J. Sham, Phys. Rev. {\bf 140}, A1133 (1965).
\bibitem{mochan85}%
%Surface Local-Field Effect,
W. Luis Moch\'an and R. G. Barrera, J. de Phys. (Par\'is) {\bf 45 C5}, 207-212 (1984);
%Intrinsic surface-induced optical anisotropies of cubic crystals: 
%Local field effect,
%	W. Luis Moch\'an y Rub\'en G. Barrera,
Phys. Rev. Lett. {\bf 55}, 1192, (1985).

\end{thebibliography}
\end{document}